\documentclass[a4paper,11pt]{article}
\usepackage{pos}
\usepackage{wrapfig}
\usepackage{adjustbox}
\usepackage{amsmath}
\usepackage{todonotes}

\title{Form factors for the charm-baryon semileptonic decay $\Xi_c\to \Xi \ell \nu$ from domain-wall lattice QCD}
\ShortTitle{Form factors for the charm-baryon semileptonic decay $\Xi_c\to \Xi \ell \nu$}

\author*{Callum Farrell}
\author{Stefan Meinel}

\affiliation{Department of Physics, University of Arizona, Tucson, AZ 85721, USA}

\emailAdd{callumf@arizona.edu}
\emailAdd{smeinel@arizona.edu}

\abstract{Recent experimental progress measuring the branching fractions of the heavy-baryon semileptonic decays $\Xi_c\to \Xi \ell \nu$ has stimulated theoretical interest and motivates precise lattice calculations of the form factors. Here we present such a calculation using domain-wall fermions for the up, down, and strange quarks, and an anisotropic clover action for the charm quark. We use four ensembles generated by the RBC and UKQCD collaborations, with lattice spacings between 0.111 and 0.073 fm and pion masses ranging from 420 to 230 MeV. Our preliminary results for the form factors are larger in magnitude than previous lattice results.}

\FullConference{The 40th International Symposium on Lattice Field Theory (Lattice 2023)\\
July 31st - August 4th, 2023\\
Fermi National Accelerator Laboratory\\}

\begin{document}
\maketitle

\section{Introduction}
\label{intro}
Although the majority of precision studies of weak decays of heavy quarks have focused on the mesonic sector, semileptonic decays of heavy baryons, e.g. $\Lambda_b \to \Lambda \ell^+ \ell^-$ \cite{Detmold:2016pkz}, $\Lambda_b \to p \ell^- \overline{\nu}_\ell$, $\Lambda_b \to \Lambda_c \ell^- \overline{\nu}_\ell$ \cite{Detmold:2015aaa}, and $\Lambda_c \to p\ell^+\ell^-$ \cite{Meinel:2017ggx}, have provided new and fruitful perspectives from which to examine flavor physics.

Recently, the BESIII Collaboration performed precise measurements of the $\Lambda^+_c \rightarrow \Lambda e^+ \nu_{e}$ and $\Lambda^+_c \rightarrow \Lambda \mu^+ \nu_{\mu}$ branching fractions and angular observables \cite{BESIII:2022ysa,BESIII:2023vfi}, enabling a detailed comparison with the lattice-QCD predictions from Refs.~\cite{Meinel:2016dqj,Meinel:2021grq}. While the $q^2$-differential decay rates are in reasonable agreement with the predictions, some tensions were seen in the slopes of individual form factors.
It is natural to extend such studies to the $SU(3)$ partner process $\Xi_c\to \Xi \ell^+ \nu_\ell$. In 2021, both Belle and ALICE performed measurements of the relative branching ratio \cite{Belle:2021crz,ALICE:2021bli} 
\begin{align*}
    \frac{B(\Xi_c^0 \rightarrow \Xi^- e^+ \nu_{e})}{B(\Xi_c^0 \rightarrow \Xi^- \pi^+)}.
\end{align*}
Combining these results with a 2018 Belle measurement of the $B(\Xi_c^0 \rightarrow \Xi^- \pi^+)$ normalization mode \cite{Belle:2018kzz} gives the following experimental results for the absolute branching fractions:
\begin{align}
    B_{\text{Belle}}(\Xi_c^0 \rightarrow \Xi^- e^+ \nu_{e}) =& (1.31 \pm 0.04 \pm 0.07 \pm 0.38 ) \%, \\
    B_{\text{ALICE}}(\Xi_c^0 \rightarrow \Xi^- e^+ \nu_{e}) =& (2.48 \pm 0.25 \pm 0.40 \pm 0.72) \%. 
\end{align} 
Here the first and second uncertainties are statistical and systematic, respectively, while the third is propagated from the uncertainty in the $\Xi_c^0 \rightarrow \Xi^- \pi^+$ normalization mode. The 2023 Review of Particle Physics (using only the Belle measurement as input) \cite{Workman:2022ynf} gives a somewhat different value, obtained from a fit including other modes, of
\begin{align}
    B_{\text{PDG}}(\Xi_c^0 \rightarrow \Xi^- e^+ \nu_{e}) =& (1.04 \pm 0.24) \%. 
\end{align} 

In the decade preceding these experimental results, a variety of different model-dependent theoretical predictions of the branching ratio were published, with the results summarized in Table \ref{modeldep}. These model-dependent estimations trend noticeably higher than the experimental measurements discussed above. Note that these predictions do not all use the same value of the $\Xi_c$ lifetime. Because of a precise LHCb measurement in 2019 \cite{LHCb:2019ldj}, the $\Xi_c^0$ lifetime was updated from $\tau_{\Xi^0_c} = (112 \pm 13) \text{ fs}$ to $\tau_{\Xi^0_c}= (154.5 \pm 2.5) \text{ fs}$. Another LHCb measurement published in 2021 then lead to the average $\tau_{\Xi^0_c}= (152.0 \pm 2.0) \text{ fs}$ \cite{LHCb:2021vll}. In Table \ref{modeldep}, the calculations marked with a $(*)$ used the pre-2019 value of $\tau_{\Xi^0_c}$.

 \begin{table}[]
    \centering
    \begin{adjustbox}{max width = \textwidth}
        \begin{tabular}[size]{l c c c} 
            \hline\hline \\[-2.2ex]
             & Method & $B(\Xi_c^0 \rightarrow \Xi^- e^+ \nu_{e})$ \\
            \hline \\[-2.2ex]
            Geng, Liu, and Tsai, 2021  \cite{Geng:2020gjh} & light-front quark model  & $(3.49 \pm 0.95) \%$  \\ 
            Zhao, 2021 \cite{Zhao:2021sje} & QCD sum rules  & $(3.4 \pm 1.7)\%$  \\
            Faustov and Galkin, 2019 \cite{Faustov:2019ddj} & rel. quark model  & $2.38\: \%^*$  \\  
            Geng et al., 2019 \cite{Geng:2019bfz} &  SU(3) & $(3.0 \pm 0.3)\%^*$  \\
            Zhao, 2018 \cite{Zhao:2018zcb} & light-front quark model  & $1.35\:\%^*$  \\
            Geng et al., 2018 \cite{Geng:2018plk} & SU(3)  &  $(4.87 \pm 1.74) \%^*$ \\
            Geng et al., 2017 \cite{Geng:2017mxn} & SU(3)  &  $(11.9 \pm 1.6)\%^*$ \\
            Azizi, Sarac, and Sundu, 2012 \cite{Azizi:2011mw} & light-cone QCD sum rules  & $(7.26 \pm 2.54)\%^*$  \\
            Liu and Huang, 2010 \cite{Liu:2010bh} & QCD sum rules & $2.4\:\%^*$  \\
            \hline\hline
        \end{tabular}
    \end{adjustbox}
    \caption{Recent model-dependent theoretical predictions of $B(\Xi_c^0 \rightarrow \Xi^- e^+ \nu_{e})$. The calculations denoted with a (*) used an outdated value of  $\tau_{\Xi^0_c}$, as explained in the main text.}
    \label{modeldep}
\end{table} 

Since the publication of the experimental results, a significant amount of theoretical interest in this decay mode has centered around the expectations of flavor $SU(3)$ symmetry. In Refs.~\cite{Zhang:2021oja, He:2021qnc, Geng:2022yxb}, the authors argue that the experimental measurements of $B(\Xi_c^0 \rightarrow \Xi^- e^+ \nu_{e})$ are considerably smaller than would be suggested by $SU(3)$ considerations based on $B(\Lambda^+_c \rightarrow \Lambda e^+ \nu_{e})$. Reference~\cite{Geng:2022yxb} suggested this could be resolved by a large $\Xi_c - \Xi'_c$ mixing angle, but subsequent lattice calculations \cite{Liu:2023feb, Liu:2023pwr} found a negligible mixing angle, in accordance with expectations from heavy quark theory. 

The first lattice calculation of the $\Xi_c\to \Xi \ell^+ \nu_\ell$ form factors was performed by Zhang et al. \cite{Zhang:2021oja}, which gave the Standard-Model prediction 
\begin{align}
    B_{\text{Lattice}}(\Xi_c^0 \rightarrow &\Xi^- e^+ \nu_{e})= (2.38 \pm 0.30 \pm 0.32 ) \%
\end{align}
for the branching ratio. The form factors obtained in that work were in fact already used in the Monte-Carlo event generation for the Belle measurement in Ref.~\cite{Belle:2021crz}, and contributed approximately 3\% to the overall systematic uncertainty. The calculation in Ref.~\cite{Zhang:2021oja} used two ensembles of lattice gauge configurations with pion masses of 290 and 300 MeV, and a clover action for all of the fermions. 

Further lattice studies of the $\Xi_c\to \Xi \ell^+ \nu_\ell$ transition can help to pin down these (mildly) discrepant values of the branching ratio and help to test the model-dependent calculations in Table \ref{modeldep}. Furthermore, the $\Xi_c \to \Xi \ell^+ \nu_\ell$ decay mode, while not as topical for flavor physics, can provide a cross-check on the control of systematics for other lattice calculations that are more relevant in the flavor sector.

The  $\Xi_c\to \Xi \ell^+ \nu_\ell$ decay amplitude depends on six form factors that parameterize the hadronic matrix elements of the weak effective Hamiltonian. In this work, we use a helicity-based definition of the form factors \cite{Feldmann:2011xf}, which has the advantage that each form factor can be individually extracted from a single ratio of two-point and three-point correlation functions. 

The matrix elements of the vector current, $\langle \Xi(p^\prime,s^\prime) | \overline{s} \,\gamma^\mu\, c | \Xi_c(p,s) \rangle$, are parameterized by the form factors $\{f_+,f_\perp,f_0\}$, and the corresponding matrix elements of the axial-vector current are parameterized by $\{g_+,g_\perp,g_0\}$. These six form factors are each functions of $q^2$, the square of the four-momentum transferred to the lepton pair, $q=p-p'$. The explicit form of the decomposition into these form factors is given in Ref.~\cite{Feldmann:2011xf}.

\section{Lattice Setup}
The 2+1 flavor domain-wall fermion ensembles used in this work were generated by the RBC and UKQCD collaborations \cite{RBC:2010qam, RBC:2014ntl, Boyle:2018knm}. The light quarks are implemented with identical parameters for both the sea and valance quarks, but the valance strange quark masses are tuned to the physical value and differ slightly from the sea quark masses \cite{RBC:2014ntl, Boyle:2018knm}. The relevant parameters are listed in Table \ref{ensembles}. All of the ensembles utilized an Iwasaki action for the gauge fields. The ``F1M'' ensemble implements the fermion fields with a M\"obius domain-wall action instead of the Shamir domain-wall action used for the other three ensembles \cite{Boyle:2018knm}. 
\begin{table}[]
    \centering
    \begin{adjustbox}{max width = \textwidth}
    \begin{tabular}[size]{c c c c c c c c c } 
        \hline\hline \\[-2.2ex]
        Label & $N_s^3 \times N_t \times N_5 $ & $a\text{ [fm]}$ & $am_{u,d}$ &  $am_{s}^{\text{(sea)}}$ &  $am_{s}^{\text{(val)}}$ & $m_\pi \text{ [MeV]}$ &  $N_{\text{ex}}$ & $N_\text{sl}$ \\
        \hline \\[-2.2ex]
        C01 & $24^3 \times 64 \times 16$ & $\approx$ 0.111 & 0.01 & 0.04 & 0.0323 & $\approx$ 420 & 283 & 2264 \\ 
        C005 & $24^3 \times 64 \times 16 $ & $\approx$ 0.111 & 0.005 & 0.04 & 0.0323 & $\approx$ 340 & 311 & 2488  \\
        F004 & $32^3 \times 64 \times 16 $ & $\approx$ 0.083 & 0.004 & 0.03 & 0.0248 & $\approx$ 300 & 251 & 2008 \\
        F1M & $48^3 \times 96 \times 12 $ & $\approx$ 0.073 & 0.002144 & 0.02144 & 0.02217 & $\approx$ 230 & 113 & 1808 \\
        \hline\hline
    \end{tabular}
    \end{adjustbox}
    \caption{Parameters of the lattice actions and propagators used in this calculation. Here $N_\text{ex}$ and $N_\text{sl}$ are the numbers of ``exact'' and ``sloppy'' samples used in the all-mode-averaging procedure described in Refs.~\cite{Blum:2012uh, Shintani:2014vja}.}
    \label{ensembles}
\end{table} 
The charm quark was implemented with an anisotropic clover action \cite{El-Khadra:1996wdx, Aoki:2001ra, Aoki:2003dg, Lin:2006ur},
tuned to remove heavy-quark discretization errors proportional to all powers of $am_Q$. The explicit form of the action, the tuning process, and the parameter values are discussed in Ref.~\cite{Meinel:2023wyg}.
We use $\mathcal{O}(a)$-improved vector and axial-vector currents of the same form as in Ref.~\cite{Detmold:2015aaa}. The renormalization is done using a mostly nonperturbative method, with $Z_V^{cc}$ and $Z_V^{ss}$ given in Ref.~\cite{Meinel:2023wyg} and residual matching factors $\rho_\Gamma$ given in Ref.~\cite{Meinel:2016dqj}.

\section{Lattice Calculation and Preliminary Results}

\begin{figure}
    \centering
    \includegraphics[scale=.35]{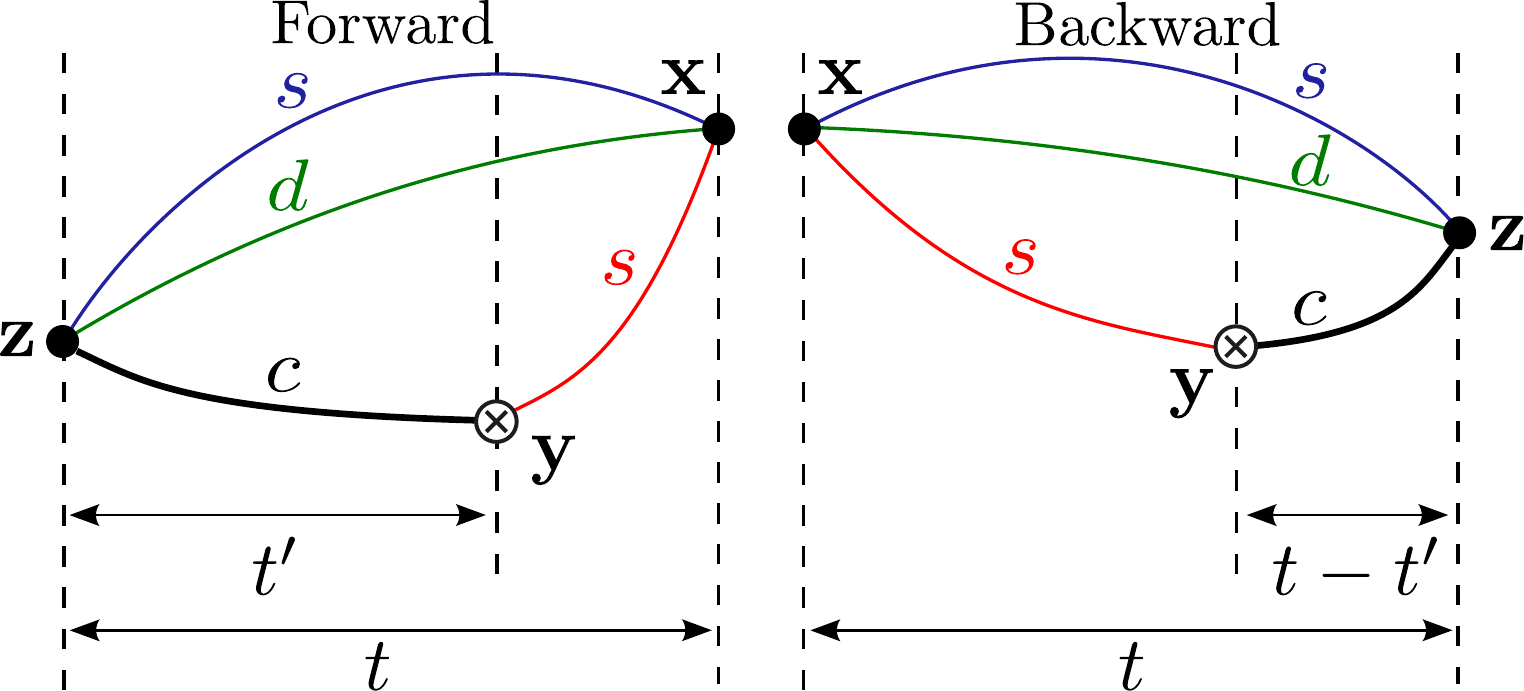}
    \caption{Diagrammatic representation of the forward and backward three-point correlation functions. The light- and strange-quark propagators have Gaussian-smeared sources at the point $(x_0,\mathbf{x})$, and the heavy-quark propagators are computed via the sequential-source method.}
    \label{3ptcorr}
\end{figure}

\begin{figure}[t]
    \centering
    \includegraphics[scale=.55]{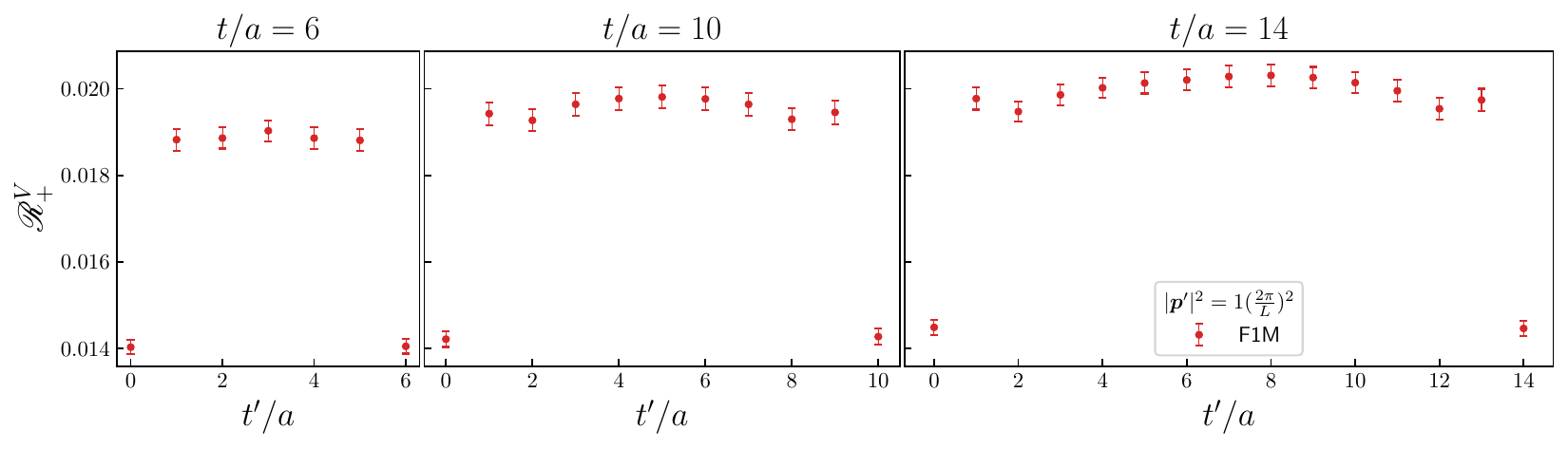}
    \caption{Example numerical values of $\mathscr{R)}_{+}^V$  at $|\mathbf{p}'|^2 = 1(2\pi/L)^2$, plotted as a function of the current-insertion time $t'$ for three different source-sink separations.}
    \label{t'}
\end{figure}

We extract the form factors from ratios of three-point and two-point correlation functions. Since the $\Xi_c$ and the $\Xi$ both have $J^P = \frac{1}{2}^+$, we use the interpolating field operators 
\begin{equation}
    \label{interpolating_fields}
     \begin{aligned}[t]
        \Xi_{c{\alpha}} &= \epsilon_{abc} (C\gamma_5)_{\beta\gamma} u^a_\beta s^b_\gamma c^c_\alpha, 
     \end{aligned}   
     \qquad 
     \begin{aligned}[t]
         \Xi_{\alpha} &= \epsilon_{abc} (C\gamma_5)_{\beta\gamma} u^a_\beta s^b_\gamma s^c_\alpha.
     \end{aligned}
\end{equation}
With these, we construct both the ``forward'' and ``backward'' three-point correlation functions
\begin{equation}
    \begin{aligned}
        C^{(3,\text{fw})}_{\delta \alpha}(\Gamma,\mathbf{p}',t,t') = \sum_{\mathbf{y},\mathbf{z}} e^{-i\mathbf{p}' \cdot (\mathbf{x}-\mathbf{y})} \Big<\Xi_{\delta}(x_0, \mathbf{x}) \:\: J^{\dagger}_\Gamma(x_0 -t + t', \mathbf{y}) \:\:  \overline{\Xi}_{c\alpha}(x_0-t,\mathbf{z}) \Big>, \\
        C^{(3,\text{bw})}_{\delta \alpha}(\Gamma,\mathbf{p}',t,t-t') = \sum_{\mathbf{y},\mathbf{z}} e^{-i\mathbf{p}' \cdot (\mathbf{y}-\mathbf{x})} \Big<\Xi_{c\delta}(x_0 + t, \mathbf{z}) \:\: J_\Gamma(x_0 + t', \mathbf{y}) \:\:  \overline{\Xi}_\alpha(x_0,\mathbf{x}) \Big>,
    \end{aligned}
\end{equation}
which are illustrated diagrammatically in Fig.~\ref{3ptcorr}. Here, $\mathbf{p}'$ is the momentum of the final-state $\Xi$ baryon, and $t$, $t'$ are the source-sink separation and current-insertion time, respectively. The corresponding two-point correlation functions are
\begin{equation}
    \begin{aligned}
        C^{(2,\Xi,\text{fw})}_{\delta \alpha}(\textbf{p}',t) &= \sum_{\textbf{y}} e^{-i\textbf{p}' \cdot (\textbf{y}-\textbf{x})} \big<\Xi_{\delta}(x_0+t, \textbf{y}) \: \overline{\Xi}_\alpha(x_0,\textbf{x}) \big>, \\
        C^{(2,\Xi,\text{bw})}_{\delta \alpha}(\textbf{p}',t) &= \sum_{\textbf{y}} e^{-i\textbf{p}' \cdot (\textbf{x}-\textbf{y})} \big<\Xi_{\delta}(x_0, \textbf{x}) \: \overline{\Xi}_\alpha(x_0-t,\textbf{y}) \big>,\\
        C^{(2,\Xi_c,\text{fw})}_{\delta \alpha}(t) &= \sum_{\textbf{y}} \big<\Xi_{c\delta}(x_0+t, \textbf{y}) \: \overline{\Xi}_{c\alpha}(x_0,\textbf{x}) \big>, \\
        C^{(2,\Xi_c,\text{bw})}_{\delta \alpha}(t) &= \sum_{\textbf{y}} \big<\Xi_{c\delta}(x_0, \textbf{x}) \: \overline{\Xi}_{c\alpha}(x_0-t,\textbf{y}) \big>. 
    \end{aligned}
\end{equation}
Following the same procedure as Ref.~\cite{Detmold:2015aaa}, we can obtain the form factors through the construction of ratios like
\begin{align}
    \mathscr{R}_{+}^V(\mathbf{p}^\prime,t,t^\prime) &=r_\mu[(1,\mathbf{0})] \: r_\nu[(1,\mathbf{0})] \frac{ \mathrm{Tr}\Big[   C^{(3,{\rm fw})}(\mathbf{p^\prime},\:\gamma^\mu, t, t^\prime) \:    C^{(3,{\rm bw})}(\mathbf{p^\prime},\:\gamma^\nu, t, t-t^\prime)  \Big] } 
    {\mathrm{Tr}\Big[C^{(2,\Xi,{\rm av})}(\mathbf{p^\prime}, t)\Big] \mathrm{Tr}\Big[C^{(2,\Xi_c,{\rm av})}(t)\Big] } ,
    \label{ratio1}
\end{align}
which cancel the overlap factors and exponential time-dependence of the ground-state contribution. The ratio is contracted with the virtual polarization vector
\begin{align}
    r[n]= n- \frac{(q \cdot n)}{q^2}q
\end{align}
to project to definite helicity. Here, $q$ is the momentum transfer to the lepton pair. From the ratio in Eq.~(\ref{ratio1}), we construct the quantity
\begin{align}
   \label{eq:Rfplus} R_{f_+}(|\mathbf{p}^\prime|, t)      =& \frac{2\, q^2 } {(E_\Xi-m_\Xi)(m_{\Xi_c}+m_\Xi)} \sqrt{\frac{ E_\Xi  }{ E_\Xi+m_\Xi } \mathscr{R}_{+}^V(|\mathbf{p}^\prime|, t, t/2)} \\
    =& f_+ + ({\rm excited\text{-}state\:\:contributions}), \nonumber
\end{align}
which is equal to the form factor at sufficiently large Euclidean times. A more complete discussion of this derivation and the ratios that give the other vector and axial-vector form factors can be found in Ref.~\cite{Detmold:2015aaa}. The $t^\prime$ dependence of $\mathscr{R)}_{+}^V$ is mild, as demonstrated in Fig.~\ref{t'}. 
In Eq.~(\ref{eq:Rfplus}), we evaluate $\mathscr{R)}_{+}^V$ at $t'=t/2$, where the excited-state contamination at each $t$ is minimal.

\begin{figure}[b]
    \centering
    \hspace{-1.55cm}
    \includegraphics[width=1\textwidth]{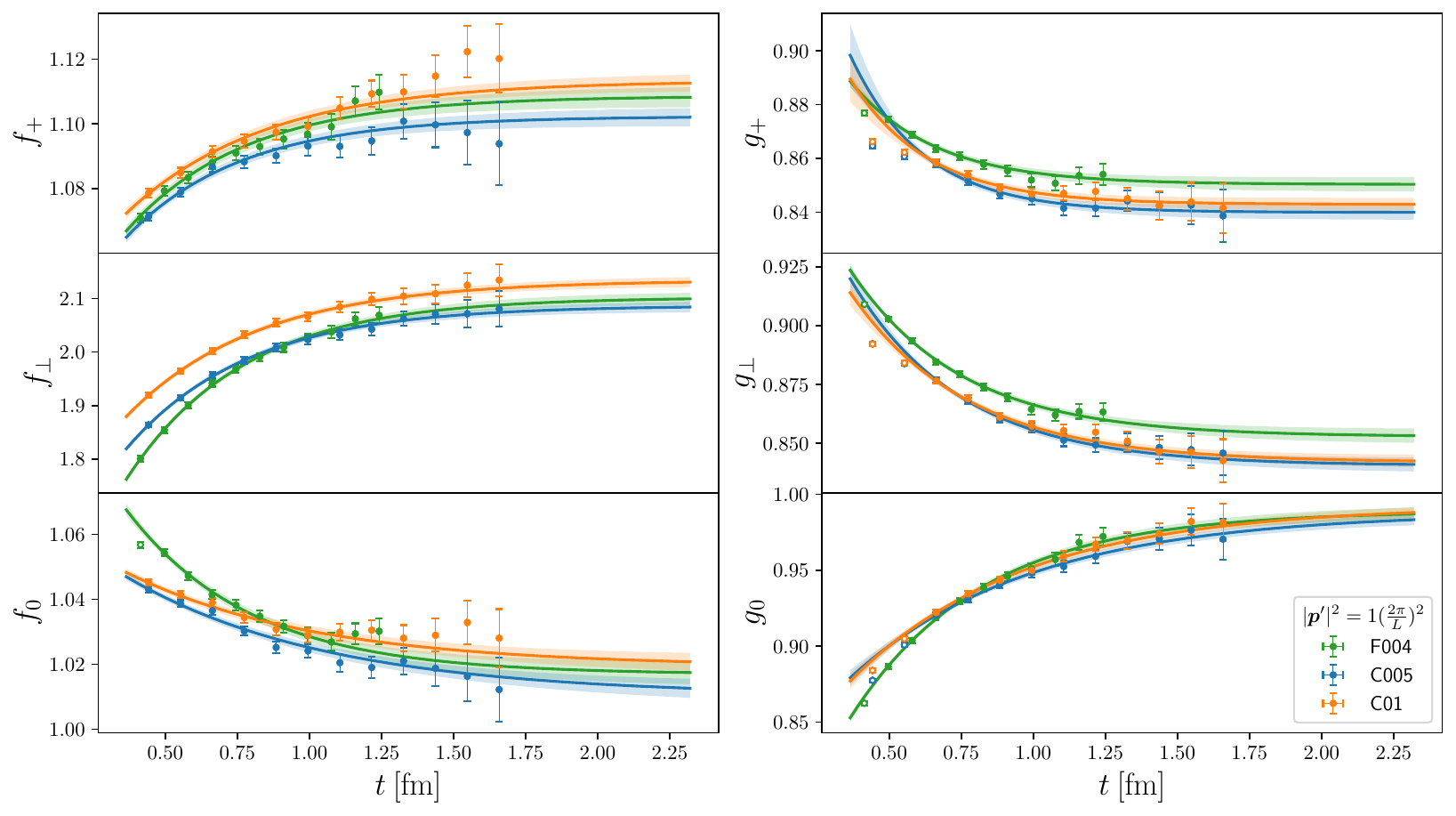}
    \caption{Example fits of $R_{f}(|\mathbf{p}^\prime|, t)$ as a function of the source-sink separation $t$. These approach the ground-state form factors at large $t$. Here we show results for both the vector (left) and axial-vector (right) form factors from the F004, C01, and C005 ensembles at $|\mathbf{p}^\prime|^2 = 1(2\pi/L)^2$. }
    \label{fits}
\end{figure}

To extract the values of the form factors at infinite source-sink separation, we use the fit functions
\begin{align}
    R_{f,i,n}(t) = f_{i,n} + A_{f,i,n} \: e^{-\delta_{f,i,n}\:t}, \hspace{2ex} \delta_{f,i,n}=\delta_{\rm min} + e^{\,l_{f,i,n}}\:\:{\rm GeV}.
\end{align}
The constant term $f_{i,n}$ is the value of the form factor at infinite source-sink separation, and the exponential term captures the dominant excited-state contamination. The index $f$ labels the different form factors, $i$ labels the different ensembles, and $n$ labels the value of the final state momentum $|\mathbf{p}'|^2=n (2\pi/L)^2$. Here, the minimum energy gap $\delta_{\rm min}$ is set to 100 MeV (smaller than any expected gap in spectrum), which helps stabilize the fits. The ensembles C01, C005 and F004 have equal spatial extent $L$ (within uncertainties) and therefore share the same values of $|\mathbf{p}'|$. We use that fact to perform a coupled fit across these three ensembles for all the vector ($f=f_+,f_\perp, f_0$) or all the axial-vector ($f=g_+,g_\perp, g_0$) form factors at a given momentum $n$ (the F1M data are fit separately from the others). The fit is coupled through the inclusion of several Gaussian priors to the $\chi^2$ function. Since the momenta are equal (within uncertainties) across the different ensembles, we anticipate that the energy gap parameters $l_{f,i,n}$ should also be similar across the different ensembles, allowing for slight variations due to the differing lattice spacings and quark masses. As described in Ref.~\cite{Detmold:2015aaa}, the priors constrain the variations of the $l_{f,i,n}$ parameters across the different ensembles to be not unnaturally large. A good $\chi^2/{\rm d.o.f.}$ was achieved by first including all values of $t$ in the fit, and then steadily increasing the minimum value of $t$ included. Examples of the fits are shown in Fig.~\ref{fits}.

\begin{figure}[!htb]
    \centering
    \includegraphics[width=\textwidth]{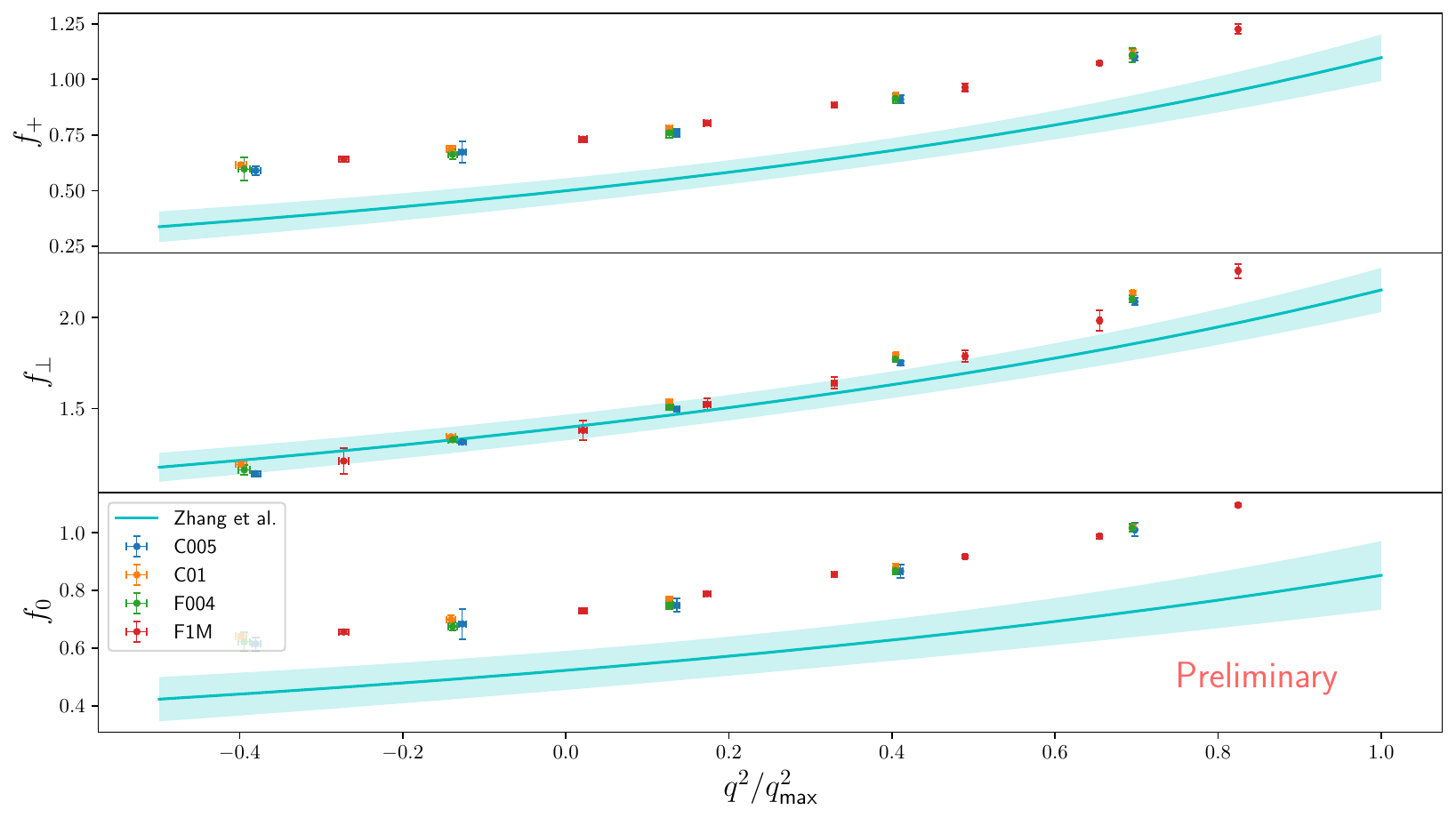}
    \caption{Comparison of our preliminary lattice results for the $\Xi_c\to\Xi$ vector form factors from the four different ensembles to the continuum-extrapolated results for these form factors (cyan curves with $1\sigma$ error bands) from the lattice calculation of Zhang et al.~\cite{Zhang:2021oja}.}
    \label{vector_ff}
\end{figure}

\begin{figure}[!htb]
    \centering
    \includegraphics[width=\textwidth]{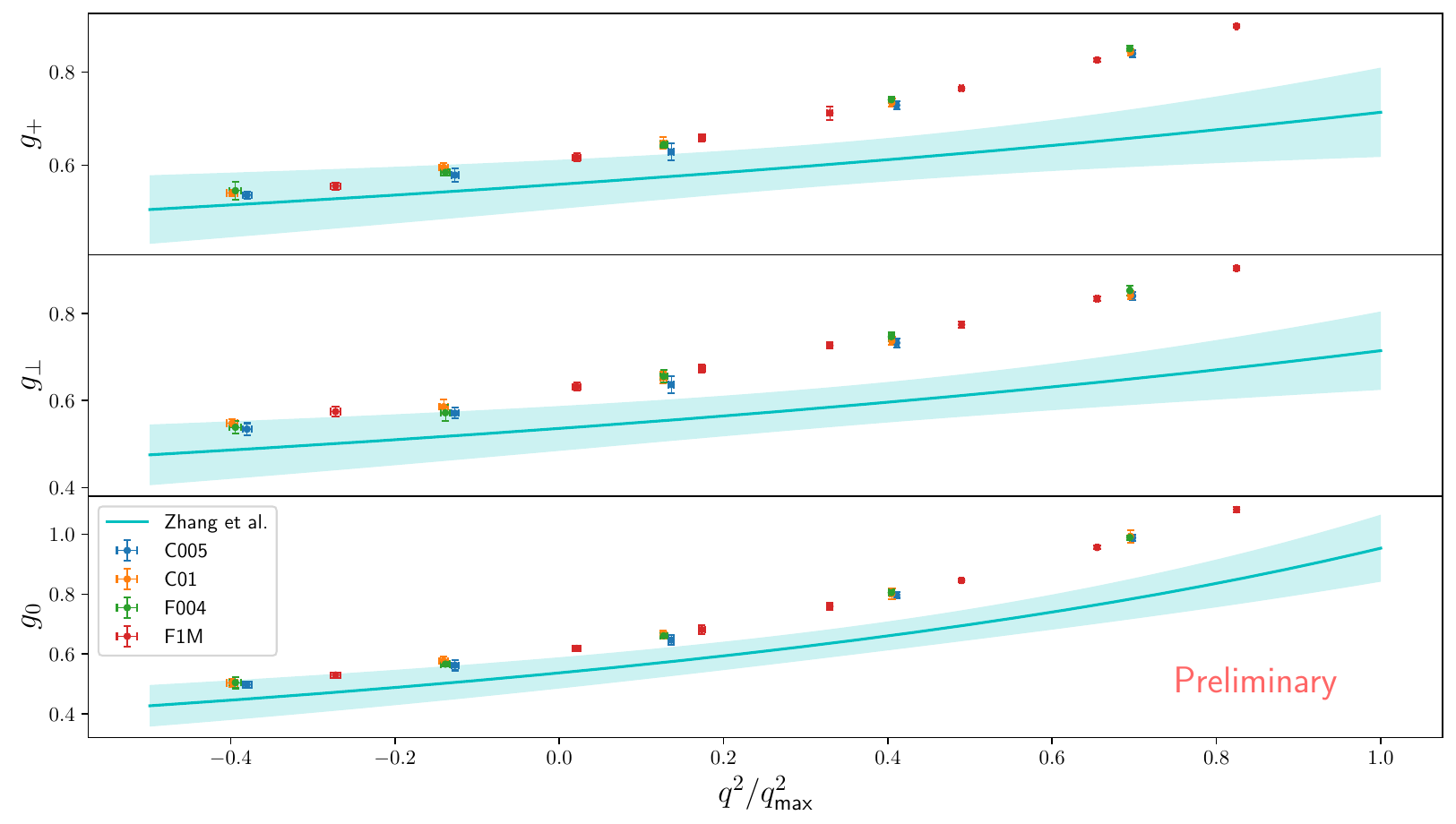}
    \caption{Like Fig.~\protect\ref{vector_ff}, but for the axial-vector form factors.}
    \label{axial_vector_ff}
\end{figure}

We have not yet performed a chiral-continuum extrapolation, so in Figs.~\ref{vector_ff} and \ref{axial_vector_ff} we show our data points for the ground-state form factors from the different ensembles in comparison to the continuum-extrapolated form factors from Ref.~\cite{Zhang:2021oja}. Our results include preliminary estimates of the systematic uncertainties from the choice of $t_{\rm min}$ in the fits to $ R_{f,i,n}(t)$, equal to the shifts in the central values of the form factors when increasing $t_{\rm min}$ by one step. These estimates are presently computed individually for each point, and we find that the sizes of these estimates can vary substantially from point to point.

We observe only mild dependence on the pion mass and lattice spacing, indicating that our chiral-continuum extrapolated form factors will, most likely, lie close to our data points.
This implies that our final results for the form factors, especially in the high-$q^2$ region, will likely be larger than those of Ref.~\cite{Zhang:2021oja}. Because each of these form factors contributes additively to the decay rate, we expect our Standard-Model prediction for the branching ratio $B(\Xi_c^0 \rightarrow \Xi^- e^+ \nu_{e})$ to be higher than the previous lattice prediction in Ref.~\cite{Zhang:2021oja}, and also higher than the experimental values in Refs.~\cite{Belle:2021crz, ALICE:2021bli}. Our prediction will likely be closer to some of the model-dependent calculations discussed in Section \ref{intro}. If this higher value of the branching ratio is confirmed, it will substantially mitigate the apparent larger-than-expected deviations from $SU(3)$ symmetry discussed in Refs.~\cite{Zhang:2021oja, He:2021qnc, Geng:2022yxb}. 

\vspace{1cm}

\section*{Acknowledgments}

We thank the RBC and UKQCD collaborations for providing the gauge-field ensembles. CF and SM are supported by the U.S. Department of Energy, Office of Science, Office of High Energy Physics under Award Number D{E-S}{C0}009913. We carried out the computations for this work on facilities at the National Energy Research Scientific Computing Center, a DOE Office of Science User Facility supported by the Office of Science of the U.S. Department of Energy under Contract No.~DE-AC02-05CH1123, and on facilities of the Extreme Science and Engineering Discovery Environment (XSEDE), which was supported by National Science Foundation grant number ACI-1548562. We used Chroma \cite{Edwards:2004sx,Chroma}, QLUA \cite{QLUA}, MDWF \cite{MDWF}, and related USQCD software \cite{USQCD}.

\providecommand{\href}[2]{#2}\begingroup\raggedright\endgroup
\end{document}